\documentclass[12pt, prd, showpacs]{revtex4}
\usepackage{amssymb}
\usepackage{amsmath}

\input{tcilatex}

\begin{document}

\title{Can a nonextremal black hole be a particle accelerator?}
\author{O. B. Zaslavskii}
\affiliation{Department of Physics and Technology, Kharkov V.N. Karazin National
University, 4 Svoboda Square, Kharkov 61022, Ukraine}
\affiliation{Institute of Mathematics and Mechanics, Kazan Federal University, 18
Kremlyovskaya St., Kazan 420008, Russia}
\email{zaslav@ukr.net }

\begin{abstract}
We consider particle collisions in the background of a nonextremal black
hole. Two particles fall from infinity, particle 1 is fine-tuned (critical),
collision occurs in its turning point. The first example is the
Reissner-Nordstr\"{o}m (RN) one. If the energy at infinity $E_{1}$ is big
enough, the turning point is close to the horizon. Then, we derive a simple
formula according to which $E_{c.m.}\sim E_{1}\kappa ^{-1/2}$, where $\kappa 
$ is a surface gravity. Thus significant growth of $E_{c.m.}$ is possible if
(i) particle 1 is ultrarelativistic (if both particles are
ultrarelativistic, this gives no gain as compared to collisions in flat
space-time), (ii) a black hole is near-extremal (small $\kappa $). In the
scenario of multiple collisions the energy $E_{c.m.}$ is finite in each
individual collision. However, it can grow in subsequent collisions,
provided new near-critical particles are heavy enough. For neutral rotating
black holes, in case (i) a turning point remains far from the horizon but
large $E_{c.m.}$ is still possible. Case (ii) is similar to that for
collisions in the RN metric. We develop a general theoretical scheme, direct
astrophysical applications can be a next step to be studied.
\end{abstract}

\keywords{particle collision, high energy collisions, centre of mass frame}
\pacs{04.70.Bw, 97.60.Lf }
\maketitle

\section{Introduction}

During last decade, a lot of efforts was devoted to high energy processes
near black holes. A large series of papers was triggered by the observation
made by Ba\~{n}ados, Silk and West \cite{ban}. They noticed that if two
particles fall towards a black hole and one of particles is fine-tuned
(critical), the energy $E_{c.m.}$ can grow unbounded (this is the so-called
BSW effect, after the names of its authors). It is important that a black
hole was supposed to be extremal. Moreover, in \cite{berti} the
impossibility of astrophysical black holes to be exactly extremal was
considered as an obstacle to the realization of this effect. This was
repeated many times in subsequent works. The main objection against the
counterpart of the BSW effect for nonextremal black holes consists in that
the critical particle cannot approach the horizon in this case. But if both
particles are not fine-tuned (they are called "usual"), $E_{c.m.}$ remains
modest.

Meanwhile, in \cite{gp} an important observation was made. Let one particle
be not exactly critical but, instead, near-critical. Then, one can adjust
the deviation from the critical state to the proximity of the point of
collision to the horizon in such a way, that $E_{c.m.}$ becomes unbounded.
However, one difficulty remains for nonextremal black holes. The most
physically interesting situation arises when both particles fall from
infinity. This can be realized for extremal black holes. But for nonextremal
ones, the potential barrier prevents a near-critical particle from reaching
the horizon in the same manner as this happens for an exactly critical one.

To overcome this difficulty, the scenario of multiple scattering was
proposed in \cite{gp}. According to it, particles 1 and 2 come from infinity
and collide close to the horizon, creating particles 3 and 4. In doing so,
particle 3 is almost critical. Afterwards, a new particle 5 coming from
infinity collides with particle 3 producing an indefinitely large $E_{c.m.}$
However, straightforward application of the multiple scattering scenario is
not fruitful. On the first glance, one can obtain finally unbounded $%
E_{c.m.}(3,5)$ in this way (arguments in parentheses indicate particle
numbers). The problem is, however, that if particles 1 and 2 are both usual,
particle 3 cannot be near-critical. Indeed, $E_{c.m.}(1,2)=E_{c.m.}(3,4)$.
Meanwhile, it follows from general principles \cite{ban}, \cite{gp}, \cite%
{prd} that near-horizon collision of two usual (or two near-critical)
particles 1 and 2 with finite individual energies leads to bounded $%
E_{c.m.}(1,2)$ while collision between the critical and usual particles
gives unbounded $E_{c.m.}(3,4)$. Thus we have a contradiction, so particle 3
with desired properties cannot appear as a result of previous collision
between particles arrived from infinity. A special case arrises when
particle 3 is not a critical in the standard sense but simply has small
individual energy $E$ \cite{eva}. However, careful analysis shows that such
a particle cannot be obtained as a result of a precedent collision too \cite%
{sp}, so the same problem remains.

In \cite{ks} the authors categorically claimed that nonextremal black holes
cannot be accelerators, provided initial particles come from infinity and
have finite individual energies $E$. Meanwhile, details of dynamics of
collision were not taken into account in \cite{ks} and this leaves some
potential gaps and questions. After the first collision, the second one can
occur much more close to the horizon. Can it lead to unbounded $E_{c.m.}$?
The main obstacle against obtaining very high $E_{c.m.}$ is related to the
fact that the critical particle cannot overcome the potential barrier on its
way to the horizon and bounces back in the turning point. But what happens
if the turning point itself becomes closer and closer to the horizon? It was
pointed out in \cite{ks} that indefinitely large $E_{c.m.}$ entails an
indefinitely large individual energy $E$. Meanwhile, the fact that $%
E\rightarrow \infty $ is required does not destroy the value of a black hole
as a particle accelerator since one can compare $E_{c.m.}$ with a similar
quantity $\left( E_{c.m.}\right) _{\infty },$ had collision would have
occurred at flat infinity. If $\left( E_{c.m.}\right) _{\infty }$ is modest
for such collision but $E_{c.m.}\gg $ $\left( E_{c.m.}\right) _{\infty }$,
this can be considered as some kind of accelerator even despite large
initial $E$. We would also like to remind a reader that collisions with very
large $E_{c.m.}$ were found to be possible if (i) a corresponding
nonextremal black is near-extremal, (ii) this includes particles on the
circular orbits \cite{innermost}, \cite{circ}.

In the present work, we consider the result of collisions when both
particles come from infinity and collide in the turning point of the
critical particle. We discuss these effects for charged static black holes
and rotating neutral ones separately. As we will see, this leaves some
possibility of nonextremal black holes to serve as particle accelerators,
although with some reservations. In doing so, the effect is achieved at the
first collision, whereas the second collision does not bring new features,
so the scenario of multiple collisions is, typically, irrelevant in the
situations under considerations. Nonetheless, there is a special
alternative. If superheavy particles can be created in collisions, this can
significantly increase the energy gain. We develop a general scheme that
enables us to understand potential possibilities of nonextremal black holes
but refrain from concrete astrophysical applications.

One reservation is in order. In papers \cite{pir1} - \cite{pir3}
indefinitely large $E_{c.m.}$ was obtained irrespective of whether the
horizon is extremal or nonextremal. Moreover, fine-tuning of a particle was
not required there. However, head-on collisions described by the first line
in eq. (2.57) of \cite{pir3} correspond to white holes (with one of
particles moving away from the horizon) rather to black holes (when both
particles move to the horizon). Such a scenario is possible but it is beyond
of scope of our work.

In what follows, we use the geometric system of units in which fundamental
constants $G=c=1$.

\section{Equations of motion}

We begin with the spherically symmetric case since it is rather simple and
admits a number of exact results. Let us consider the black hole metric%
\begin{equation}
ds^{2}=-dt^{2}f+\frac{dr^{2}}{f}+r^{2}d\omega ^{2}\,\text{,}  \label{sph}
\end{equation}%
where $d\omega ^{2}=d\theta ^{2}+\sin ^{2}\theta d\phi ^{2}$, $f=f(r).$ For
the Reissner-Nordstr\"{o}m (RN) metric, 
\begin{equation}
f=1-\frac{2M}{r}+\frac{Q^{2}}{r^{2}}=(1-\frac{r_{+}}{r})(1-\frac{r_{-}}{r})%
\text{,}
\end{equation}%
where $M$ is the mass, $Q$ being the electric charge of a nonextremal black
hole. Here, $r_{+}=M+\sqrt{M^{2}-Q^{2}}$, is the event horizon radius, $%
r_{-}=M-\sqrt{M^{2}-Q^{2}\text{ }}$is the Cauchy horizon radius, $M>Q$, $%
r_{+}>r_{-}$.

The electric potential equals%
\begin{equation}
\varphi =\frac{Q}{r}\text{.}  \label{pot}
\end{equation}

If a particle with the mass $m$ and electric charge $q$ moves in this
background and other external forces are absent, the equations of motion
give us%
\begin{equation}
m\dot{t}=\frac{X}{f}\text{,}  \label{t}
\end{equation}%
\begin{equation}
m\dot{\phi}=\frac{L}{r^{2}}\text{,}
\end{equation}%
\begin{equation}
X=E-q\varphi =E-\frac{qQ}{r}\text{,}  \label{x}
\end{equation}%
\begin{equation}
m\dot{r}=\sigma P\text{, }P=\sqrt{U}\text{, }U=X^{2}-f\tilde{m}^{2}\text{,}
\label{r}
\end{equation}%
\begin{equation}
\tilde{m}^{2}=m^{2}+\frac{L^{2}}{r^{2}},
\end{equation}%
$E$ is the energy, $L$ being the angular momentum, dot denotes derivative
with respect to the proper time, $\sigma =\pm 1$. The forward-in-time
condition $\dot{t}>0$ entails%
\begin{equation}
X\geq 0\text{.}  \label{ftw}
\end{equation}

We use the standard classification. If $X_{H}>0$ is separated from zero, we
call a particle usual.\ If $X_{H}=0$, it is called critical. If $X_{H}=O(%
\sqrt{f})$ near the horizon is small, it is called near-critical. Here, $%
X_{H}$ is the value of $X$ on the horizon.

\section{Particle collisions}

Let particles 1 and 2 collide. One can define the energy in the center of
mass frame $E_{c.m.}$ according to the relation 
\begin{equation}
E_{c.m.}^{2}=-(m_{1}u_{1\mu }+m_{2}u_{2\mu })(m_{1}u_{1}^{\mu
}+m_{2}u_{2}^{\mu })=m_{1}^{2}+m_{2}^{2}+2m_{1}m_{2}\gamma \text{,}
\label{ecm}
\end{equation}%
where $\gamma =-u_{1\mu }u^{2\mu }$ is the Lorentz factor of relative
motion, $u^{\mu }$ is the four-velocity. We consider pure radial motion of
particles in the RN background, so $L_{1}=L_{2}=0$. From equations of motion
(\ref{t}) - (\ref{r}) one finds%
\begin{equation}
m_{1}m_{2}\gamma =\frac{X_{1}X_{2}-P_{1}P_{2}}{f}\text{,}  \label{ga}
\end{equation}%
where we assumed that both particle move towards a black hole, so $\sigma
_{1}=\sigma _{2}=-1$. In particular, if collision occurs in the turning
point for one of particles (say, particle 1),%
\begin{equation}
m_{1}m_{2}\gamma =\frac{X_{1}X_{2}}{f}\text{.}  \label{tp}
\end{equation}

To simplify formulas, we assume that particle 2 is neutral. This also
enables us to avoid the question about the direct electric interaction
between particles.

In what follows, we also assume for simplicity that $m_{1}=m_{2}\equiv m$.
Then, for collision in the turning point where $P_{1}=0$, eqs. (\ref{ecm}), (%
\ref{ga}) give us 
\begin{equation}
E_{c.m.}^{2}=2m^{2}+\frac{2X_{1}E_{2}}{f},  \label{e12}
\end{equation}%
where the right hand side is taken in the turning point.

\section{Flat space-time}

Before discussion of collisions in the RN metric, it is instructive to list
the main formulas for the flat space-time. They are quite trivial by
themselves, but in what follows we will need to compare with them the
results of collision in the black hole background to check, whether
collision in the turning point gives some enhancement as compared to the
collision at infinity.

If $E_{1}\sim E_{2}\sim m\,\ $it is obvious that $E_{c.m.}\sim m$ as well.
If $E_{2}=m$,

\begin{equation}
\left( E_{c.m.}^{2}\right) _{flat}=2m^{2}+2E_{1}m.  \label{e2f}
\end{equation}

Thus if $E_{1}$ grows, $\left( E_{c.m.}^{2}\right) _{flat}$ grows as well.

If $E_{1}=E_{2}=E\gg m$, it follows from (\ref{ecm}) and (\ref{ga}) with $%
f=1 $ that%
\begin{equation}
\left( E_{c.m.}^{2}\right) _{flat}\approx 4m^{2}  \label{minf}
\end{equation}%
is finite.

\section{Allowed zone of motion}

Now, we return to the RN metric. The motion is possible where $U\geq 0$.
This condition gives us

\begin{equation}
\left( E-\frac{qQ}{r}\right) ^{2}\geq (m^{2}+\frac{L^{2}}{r^{2}})(1-\frac{2M%
}{r}+\frac{Q^{2}}{r^{2}})\text{.}
\end{equation}

We assume that $Qq>0$ (say, $Q>0$, $q>0$) since it is this case that
potentially gives us unbounded $E_{c.m.}$ \cite{jl}. In the turning points $%
U=0$. If $L=0$, we can find the turning point analytically: 
\begin{equation}
r_{1,2}=\frac{1}{\varepsilon ^{2}-1}(\varepsilon \tilde{q}Q-M\pm \sqrt{D})%
\text{,}
\end{equation}%
where $\varepsilon =\frac{E}{m}$, $\tilde{q}=\frac{q}{m}$, $r_{1}\leq r_{2}$%
. 
\begin{equation}
D=Q^{2}(\tilde{q}^{2}+\varepsilon ^{2}-1)-2M\varepsilon \tilde{q}Q+M^{2}.
\end{equation}

As a particle falls from infinity, where $E\geq m$, we have $\varepsilon
^{2}\geq 1$. Turning points outside the horizon are absent if $D<0$ or 
\begin{equation}
r_{2}<r_{+}.  \label{r2h}
\end{equation}%
In what follows we will consider the case when particle 1 is critical and
particle 2 is neutral, $q_{2}=0$. This means that the turning point $r_{1}$
is absent for particle 2,%
\begin{equation}
r_{2}=\frac{\sqrt{D}-M}{\varepsilon ^{2}-1}\text{,}
\end{equation}%
where 
\begin{equation}
D=M^{2}+Q^{2}(\varepsilon ^{2}-1)\text{.}
\end{equation}

It is easy to check that (\ref{r2h}) is satisfied, so the point $r_{2}$ is
absent too. Thus particle 2 comes from infinity and reaches the horizon.

\section{Critical particle}

For the critical particle, $X_{H}=0$, so we have from (\ref{x}) that

\begin{equation}
E=\frac{qQ}{r_{+}}\text{,}
\end{equation}

\begin{equation}
D=M^{2}-Q^{2}\text{,}
\end{equation}

\begin{equation}
r_{1}=r_{+}\text{,}  \label{r1c}
\end{equation}%
\begin{equation}
r_{2}=r_{+}+\frac{2\sqrt{M^{2}-Q^{2}}}{\varepsilon ^{2}-1}=r_{+}(1+\frac{%
2\kappa r_{+}}{\varepsilon ^{2}-1})\text{,}  \label{r2}
\end{equation}%
\begin{equation}
X(r_{2})=E(1-\frac{r_{+}}{r_{2}})=\frac{2\kappa r_{+}^{2}E}{%
r_{2}(\varepsilon ^{2}-1)}\text{.}  \label{xcr}
\end{equation}%
Here, $\kappa =\frac{1}{2}f^{\prime }(r_{+})$ is the surface gravity,%
\begin{equation}
\kappa =\frac{1}{2r_{+}}\left( 1-\frac{r_{-}}{r_{+}}\right) =\frac{\sqrt{%
M^{2}-Q^{2}}}{r_{+}^{2}}\text{.}
\end{equation}%
A special case arises if $\varepsilon =1$. Then, $\tilde{q}Q=r_{+}$, and for 
$r\rightarrow \infty $ we have $U\approx \frac{2m^{2}(M-r_{+})}{r}<0$. Such
a particle cannot move at infinity, so to avoid this case, we assume $%
\varepsilon >1$ in what follows.

\section{Collision between the critical and neutral particles}

If particles fall from infinity and collide in point $r=r_{2}$, it follows
from (\ref{e12}) that%
\begin{equation}
E_{c.m.}^{2}=2m^{2}+\frac{2E_{1}E_{2}(1-\frac{r_{+}}{r_{2}})}{f(r_{2})}\text{%
.}  \label{ecmcr}
\end{equation}

In the region $r_{+}<r<r_{2}$, motion of particle 1 is forbidden since $U$
becomes negative there.

The only hope to obtain unbounded $E_{c.m.}^{2}$ is to arrange collision
near the horizon, where $f\rightarrow 0$. So, now we examine, whether or not
this gives the unbounded $E_{c.m.}^{2}$.

The condition $f(r_{2})\ll 1$ requires $r_{2}\rightarrow r_{+}.$ As we see
it from (\ref{r2}), this happens if the second term in parentheses is small,
so%
\begin{equation}
\frac{\kappa r_{+}}{\varepsilon ^{2}-1}\ll 1\text{.}  \label{sm}
\end{equation}

There are two typical cases here.

\subsection{$\protect\kappa r_{+}=O(1)$, $\protect\varepsilon \rightarrow
\infty $}

Then, 
\begin{equation}
f(r_{2})\approx 2\kappa (r_{2}-r_{+})\approx \frac{4\kappa ^{2}r_{+}^{2}}{%
\varepsilon ^{2}}\text{,}  \label{f2}
\end{equation}%
taking into account (\ref{xcr}) we obtain%
\begin{equation}
E_{c.m.}^{2}\approx 2m^{2}+\frac{E_{1}E_{2}}{\kappa r_{+}}\text{.}
\label{ek}
\end{equation}

If $E_{2}=m$, there is no energy gain as compared to the flat case (\ref{e2f}%
). However, if not only $E_{1}\gg m$, but also $E_{2}\gg m$, collision near
the horizon is much more effective due to the factor $E_{1}E_{2}$ that is
absent in (\ref{minf}).

\subsection{$\protect\varepsilon =O(1)$, $\protect\kappa r_{+}$ $\ll 1$.}

This means that our black hole is near-extremal. Then, we must retain in the
expansion for the function $f(r)$ also the next term:%
\begin{equation}
f\approx 2\kappa (r-r_{+})+\frac{(r-r_{+})^{2}}{r_{+}^{2}}\text{,}
\end{equation}%
so%
\begin{equation}
f(r_{2})\approx 4\kappa ^{2}r_{+}^{2}\frac{\varepsilon ^{2}}{(\varepsilon
^{2}-1)^{2}}=4\kappa ^{2}r_{+}^{2}\frac{E_{1}^{2}m^{2}}{(E_{1}^{2}-m^{2})^{2}%
}\text{.}
\end{equation}%
Taking into account (\ref{ecmcr}), we obtain%
\begin{equation}
E_{c.m.}^{2}\approx \frac{E_{2}(E_{1}^{2}-m^{2})}{\kappa r_{+}E_{1}}\text{.}
\label{e2m}
\end{equation}

Independently of $E_{1}$ and $E_{2}$, we obtain formally unbounded growth
when $\kappa \rightarrow 0$.

And, the combined case $\varepsilon \gg 1$, $\kappa r_{+}$ $\ll 1$ is
possible as well. Then, (\ref{e2m}) turns into (\ref{ek}).

\section{Nonzero angular momentum}

Let us consider now the case, when $L\neq 0$ for particle 1.

Then, if particle 1 is critical, we have for it%
\begin{equation}
U=(1-\frac{r_{+}}{r})[E^{2}(1-\frac{r_{+}}{r})-(1-\frac{r_{-}}{r})\tilde{m}%
^{2}]\text{.}  \label{pt}
\end{equation}

We are interested in the situation when the turning point $r_{2}$ is close
to the horizon. Assuming%
\begin{equation}
\kappa r_{+}\frac{\tilde{m}_{1}^{2}(r_{+})}{E^{2}-\tilde{m}_{1}^{2}(r_{+})}%
\ll 1
\end{equation}%
and repeating simple calculations step by step, we obtain that if%
\begin{equation}
E\gg \tilde{m}(r_{+})  \label{em}
\end{equation}%
is satisfied, then (\ref{ek}) holds.

If $\kappa r_{+}\ll 1$,

\begin{equation}
\frac{r_{2}-r_{+}}{r_{+}}\approx \frac{\tilde{m}_{1}^{2}(r_{+})}{E^{2}-%
\tilde{m}_{1}^{2}(r_{+})}(1-\frac{r_{-}}{r_{+}})=2\kappa r_{+}\frac{\tilde{m}%
_{1}^{2}(r_{+})}{E^{2}-\tilde{m}_{1}^{2}(r_{+})}\text{.}
\end{equation}

Then we have, instead of (\ref{e2m}),%
\begin{equation}
E_{c.m.}^{2}\approx \frac{E_{2}[E_{1}^{2}-\tilde{m}^{2}(r_{+})]}{\kappa
r_{+}E_{1}}\text{,}  \label{ek1}
\end{equation}%
where now the case $E_{1}\gtrsim \tilde{m}(r_{+})$ is allowed.

The only difference as compared to the case $L=0$ consists in the fact that
the quantity $\tilde{m}(r_{+})$ appears in some formulas instead of $m$.

\section{Multiple collisions}

We see that indeed \ $E_{c.m.}^{2}$ can become large due to big $E_{1}$ or
small $\kappa $. Now, we want to elucidate, is it possible to improve the
result (\ref{ek1}) and increase $E_{c.m.}$? To this end, we consider the
following realization of multiple scattering scenario \cite{gp}. Particle 1
and 2 collide creating particles 3 and 4. We want to achieve $X_{3}$ as
small as possible. Then, in the case of success, collision between particle
3 and particle 5 coming from infinity can give large $E_{c.m.}$ Then, we can
take advantage of the results of analysis already carried out in \cite{centr}%
. Although the corresponding equations are derived in \cite{centr} for the
rotating case whereas now a black hole is static, the general formulas look
the same. For simplicity, again $m_{1}=m_{2}=m$, also $m_{3}=m_{4}$ and all
angular momenta $L_{i}=0$. Then, given parameters of particles 1 and 2, in
the point of collision (where subscript "c" will be used) one has from eqs.
(19), (20) of \cite{centr} (this can also be re-obtained directly form the
conservation laws)

\begin{equation}
\left( X_{3}\right) _{c}=\frac{1}{2}(X_{0}-P_{0}\sqrt{1-4\frac{m_{3}^{2}}{%
m_{0}^{2}}})_{c}\text{,}  \label{x1}
\end{equation}%
\begin{equation}
\left( X_{4}\right) _{c}=\frac{1}{2}(X_{0}+P_{0}\sqrt{1-4\frac{m_{3}^{2}}{%
m_{0}^{2}}})_{c}\text{,}  \label{x2}
\end{equation}%
\begin{equation}
P_{0}=\sqrt{X_{0}^{2}-m_{0}^{2}f}\text{,}
\end{equation}%
where $m_{0}=E_{c.m.}$, $X_{0}=X_{1}+X_{2}$. As before, particle 1 is
critical, particle 2 is usual. Let $\varepsilon \gg 1$ with $\kappa
r_{+}\sim 1$.

According to (\ref{f2}) and (\ref{ek}), in the point of collision near the
horizon $f=O(\frac{1}{\varepsilon ^{2}})$, $m_{0}^{2}=O(\varepsilon )$, $%
X_{0}\approx E_{2}=m$,%
\begin{equation}
P_{0}\approx X_{0}-\frac{m_{0}^{2}f}{2X_{0}}.
\end{equation}

Then,%
\begin{equation}
\left( X_{3}\right) _{c}\approx \frac{\kappa r_{+}(m^{2}+m_{3}^{2})}{%
\varepsilon m}\text{.}  \label{x3c}
\end{equation}%
Let $q_{1}=q_{3}=q$, $q_{2}=q_{4}=0$. Then, it follows from (\ref{x}), (\ref%
{r2}) that%
\begin{equation}
X_{3}=\left( X_{3}\right) _{c}+qQ(\frac{1}{r_{2}}-\frac{1}{r})\text{.}
\label{r2r}
\end{equation}%
In particular, 
\begin{equation}
X_{3}(r_{+})\approx \left( X_{3}\right) _{c}-\frac{2\kappa qQ}{\varepsilon
^{2}}.
\end{equation}%
The first term in (\ref{r2r}) has the order $\varepsilon ^{-1}$ and
dominates everywhere between $r_{+}$ and $r_{2}$ . Thus, in the main
approximation, the second term can be neglected and $X_{3}\approx \left(
X_{3}\right) _{c}$. It is convenient to make the substitution%
\begin{equation}
r-r_{+}=\frac{\left( X_{3}\right) _{c}^{2}}{2\kappa m_{3}^{2}}y\text{.}
\label{rx3}
\end{equation}%
Then, for $f\approx 2\kappa (r-r_{+})$ we have

\begin{equation}
f\approx \frac{\left( X_{3}\right) _{c}^{2}}{m_{3}^{2}}y\text{.}
\end{equation}%
Correspondingly, 
\begin{equation}
P_{3}^{2}\approx \left( X_{3}\right) _{c}^{2}-m_{3}^{2}f=\left( X_{3}\right)
_{c}^{2}(1-y)\text{.}
\end{equation}%
The collision between particles 1 and 2 occurred in the point $r=r_{2}$, for
which the corresponding value $y=y_{1}$ follows from (\ref{r2}), (\ref{rx3}%
): 
\begin{equation}
y_{1}=\frac{4m^{2}m_{3}^{2}}{(m_{3}^{2}+m^{2})^{2}}\text{.}
\end{equation}

After this collision, a new particle 3 can move either towards the horizon
with $\sigma _{3}=-1$ or reach a new turning point where $r=\tilde{r}_{2}$, $%
y=1$. In the second case, it bounces back there and moves further towards
the horizon with $\sigma _{3}=-1$.

From (\ref{rx3}) one can find a location of a new turning point:

\begin{equation}
\tilde{r}_{2}-r_{+}=\frac{\left( X_{3}\right) _{c}^{2}}{2\kappa m_{3}^{2}}=%
\frac{\kappa r_{+}^{2}(m^{2}+m_{3}^{2})^{2}}{2\varepsilon ^{2}m_{3}^{2}m^{2}}%
\text{.}
\end{equation}%
If $m_{3}=m$, this coincides with (\ref{r2}). Then, a particle either has $%
\sigma _{3}=-1$ or bounces back and changes $\sigma _{3}\,\ $to $-1$
immediately. In general,%
\begin{equation}
\frac{\tilde{r}_{2}-r_{+}}{r_{2}-r_{+}}=\frac{(m_{3}^{2}+m^{2})^{2}}{%
4m_{3}^{2}m^{2}}\geq 1\text{.}
\end{equation}

Let a usual particle 5 with the energy $E_{5}=$ $m_{5}=m$ and $q=0$ fall
from infinity, $\sigma _{5}=-1$. If collision occurs when $\sigma _{3}=+1$,
we have from (\ref{ecm}), (\ref{ga}), (\ref{x3c}) that%
\begin{equation}
E_{c.m.}^{2}\approx \frac{2E_{1}m_{3}^{2}mF_{+}(y)}{\kappa
r_{+}(m^{2}+m_{3}^{2})}\text{,}
\end{equation}%
\begin{equation}
F_{+}(y)=\frac{1+\sqrt{1-y}}{y}\text{,}
\end{equation}%
where $y\geq y_{1}$, This function is monotonically decreasing with $y$, so
it attains the maximum value at $y=y_{1}$, where%
\begin{equation}
\left( E_{c.m.}^{2}\right) _{\max }\approx \frac{2E_{1}m_{3}^{2}mF_{+}(y_{1})%
}{\kappa r_{+}(m^{2}+m_{3}^{2})}\text{.}  \label{2}
\end{equation}%
The most "profitable" case corresponds to head-on collision in the point $%
y=y_{1}$. This implies that the 2nd collision occurs in the same point as
the first one. If $m\ll m_{3}$, $y_{1}\approx \frac{4m^{2}}{m_{3}^{2}}\ll 1$%
. Then, $F_{+}(y_{1})\approx \frac{2}{y_{1}}$,%
\begin{equation}
\left( E_{c.m.}^{2}\right) _{\max }\approx \frac{E_{1}m_{3}^{2}}{\kappa
r_{+}m}\text{.}  \label{max}
\end{equation}

But, if $\kappa r_{+}=O(1)$, $E_{c.m.}^{2}$ remains limited.

If collision occurs when $\sigma _{3}=-1$, we have $\sigma _{5}\sigma _{3}=+1
$. Then, in the same manner we obtain%
\begin{equation}
E_{c.m.}^{2}\approx \frac{2E_{1}m_{3}^{2}mF_{-}(y)}{\kappa
r_{+}(m^{2}+m_{3}^{2})}\text{,}
\end{equation}%
\begin{equation}
F_{-}(y)=\frac{1-\sqrt{1-y}}{y}=\frac{1}{1+\sqrt{1-y}}\text{.}
\end{equation}%
Here, $F$ is monotonically increasing bounded function, $F(0)=\frac{1}{2}$, $%
F(1)=1$. Thus if the second collision occurs at $y=1$, the result for $%
E_{c.m.}^{2}$ is as twice as many as compared to the collision on the
horizon. This is quite similar to the observation made for the nonextremal
Kerr metric in \cite{piat} (see discussion after eq. 31 there) and
generalized in Sec. 2.2. of \cite{circ}. Thus a second collision does not
lead to unbounded $E_{c.m.}$.

We can compare $\left( E_{c.m.}^{2}\right) _{2}$ after the 2nd collision
with a similar quantity $\left( E_{c.m.}^{2}\right) _{1}$ (\ref{ek}) after
the 1st collision. Taking into account (\ref{max}), we obtain 
\begin{equation}
\frac{\left( E_{c.m.}^{2}\right) _{2}}{\left( E_{c.m.}^{2}\right) _{1}}%
\approx \frac{m_{3}^{2}}{m^{2}}\text{.}  \label{fr}
\end{equation}

If all masses have the same order $m$, there is no big gain. However, if,
say, $m_{5}=m$ but $m_{3}\gg m$, $\frac{\left( E_{c.m.}^{2}\right) _{2}}{%
\left( E_{c.m.}^{2}\right) _{1}}\gg 1$. Meanwhile, there is an upper bound
here. As $\left( E_{c.m.}\right) _{1}\geq 2m_{3}$, there is a bound%
\begin{equation}
\frac{\left( E_{c.m.}^{2}\right) _{2}}{\left( E_{c.m.}^{2}\right) _{1}}\leq 
\frac{\left( E_{c.m.}^{2}\right) _{1}}{4m^{2}}\approx \frac{E_{1}}{4m\kappa
r_{+}}\text{,}
\end{equation}%
where (\ref{ek}) with $E_{2}=m$ was used again.

One can repeat the procedure. Let a new particle 6 with $m_{6}=m$ is sent
from infinity. It collides with particle 3 and produces a new near-critical
particle 7. Repeating derivation, we obtain in the new point of collision (%
\ref{x3c}) with $m_{3}$ replaced with $m_{7}$. In eq. (\ref{fr}) $m_{3}$
should be replaced with $m_{7}$.

We can imagine a scenario in which initially a (near)critical particle 1
with $E_{1}\gg m$ is sent from infinity together with particle 2 having $%
E_{2}=m$. They collide, create a near-critical particle with $m_{3}$ that
collides with a new particle having $E=m$ and coming from infinity, etc. If,
for simplicity, all new near-critical particles have the same mass $m_{3}$
and falling particles have the same mass $m$, each time $E_{c.m.}$ can
acquire an additional factor $\left( \frac{m_{3}}{m}\right) $ that results
in $\left( \frac{m_{3}}{m}\right) ^{n}$, where $n$ is the number of
additional collisions$.$ It can be quite big, provided new near-critical
particles are heavy enough. In this scenario, a big energy $E_{1}$ is pumped
into the system but this is done only one time. It is worth noting that in
the multiple scenario suggested in \cite{gp}, only the angular momentum
changes due to collisions. Meanwhile, \ now parameters of a near-critical
particle are fixed, the effect of big $E_{c.m.}$ is achieved due to the
relation between masses of a near-critical and usual particles.

Thus if we want to obtain big $E_{c.m.}$, the near-critical particle should
be superheavy. In this sense, there is some analogy between collisions in
our scenario and collisions near extremal charged black holes. Namely, it
was shown in \cite{rn} that in the scenario denoted there OUT+, there is no
upper bound on $m_{3}$ and, instead, there is a lower bound. The similar
result was obtained in somewhat different approach in \cite{nem}. In this
sense, the collisional Penrose process with ultrahigh $E_{c.m.}$ can be
accompanied with ultrahigh $m_{3}$. Meanwhile, there is also difference
between the process under discussion in the present work and that considered
in \cite{rn}, \cite{nem}. In the extremal case, $E_{c.m.}$ and $m_{3}$ are
independent, so it can happen that $E_{c.m.}$ is ultrahigh whereas $m_{3}$
is modest (although restricted from below). Meanwhile, for nonextremal black
holes, big $m_{3}$ is a necessary condition for obtaining high $E_{c.m.}$

There is also a counterpart of the phenomenon of interrelation between $%
E_{c.m.}$ and $m_{3}$ for rotating black holes. It was found in \cite{sph}
for the Kerr metric and was generalized in \cite{max}. Then, although there
is an upper bound on $E_{c.m.}$, significant increase in $E_{c.m.}$ occurs
when a created particle is superheavy.

One additional remark is in order. As it is clear from the method of
derivation, it is not important, whether the new particle will have
parameters close to the criticality condition $X_{H}\approx 0$ due to the
compensation between $E$ and $q\varphi $ or simply it has $q=0$ and small
energy \cite{eva}, \cite{sp}.

\section{Rotating case}

It is the case of rotating black holes that we now turn to. In doing so, we
assume no electric interaction between particles and a black hole. It means
that either particles or a black hole are electrically neutral (or both a
black hole and particles). As consideration of collisions runs along the
same line, we give only brief description. The metric has the form%
\begin{equation}
ds^{2}=-N^{2}dt^{2}+g_{\phi }(d\phi -\omega dt)^{2}+\frac{dr^{2}}{A}%
+g_{\theta }d\theta ^{2}\text{,}
\end{equation}%
where for shortness $g_{\phi }\equiv g_{\phi \phi }$ and $g_{\theta }\equiv
g_{\theta \theta }$. We assume that the metric coefficients do not depend on 
$t$ and $\phi $ and possess symmetry because of which motion within the
plane $\theta =\frac{\pi }{2}$ is possible. In this plane, we can redefine
the radial coordinate to have $N^{2}=A$. Then, the equations of motion for a
free particle have the form 
\begin{equation}
m\dot{t}=\frac{X}{N^{2}}\text{,}
\end{equation}%
\begin{equation}
m\dot{\phi}=\frac{L}{g_{\phi }}+\frac{\omega X}{N^{2}}\text{,}
\end{equation}%
\begin{equation}
m\dot{r}=\sigma P\text{, }P=\sqrt{U}
\end{equation}%
with%
\begin{equation}
U=X^{2}-\tilde{m}^{2}N^{2}\text{,}  \label{U}
\end{equation}%
\begin{equation}
X=E-\omega L\text{,}
\end{equation}%
\begin{equation}
\tilde{m}^{2}=m^{2}+\frac{L^{2}}{g_{\phi }}\text{.}
\end{equation}

The main difference with respect to the RN case consists in that the
critical particle has 
\begin{equation}
L=\frac{E}{\omega _{H}}\text{,}  \label{crit}
\end{equation}%
so for it, $E$ and $L$ are not independent parameters any longer.

The rotational counterpart of eq. (\ref{e12}) for collision in the turning
point $r_{t}$ of particle 1 now reads%
\begin{equation}
E_{c.m.}^{2}=2m^{2}+\frac{2X_{1}(r_{t})X_{2}(r_{t})}{N^{2}(r_{t})}-\frac{%
2L_{1}L_{2}}{g_{\phi (}r_{t})}\text{.}  \label{erot}
\end{equation}

\subsection{Ultrarelativistic particles}

It turns out that even for ultrarelativistic particles with (\ref{em}), the
turning point does not approach the horizon. Indeed, if $N\ll 1$, $\frac{%
r-r_{+}}{r_{+}}\ll 1$ and the Taylor expansion has the form 
\begin{equation}
\omega =\omega _{H}-B_{1}(r-r_{+})+...\text{.}
\end{equation}%
where $B_{1}$ is some constant. For the critical particle,

\begin{equation}
X=\frac{E}{\omega _{H}}B_{1}(r-r_{+})+...  \label{xrot}
\end{equation}%
The first term in $U$ has the order $(r-r_{+})^{2}$ whereas the second
negative one has the order $N^{2}\sim (r-r_{+})$, so $U<0$. For the RN
metric, we were able to choose a large energy of the particle to achieve
proximity of the turning point to the horizon since large $E$ compensated
small $r-r_{+}$. But now this is impossible since the negative contribution
in $U$ has the same factor $E^{2}$ as a positive one due to condition (\ref%
{crit}).

Thus the turning point is located in some intermediate region where $N\sim 1$%
. Now, the type of particle is irrelevant at all. Let, for simplicity, both
particles be usual with $L_{1}=L_{2}=0$, so $X_{1}=E_{1}$. $X_{2}=E_{2}$.
Then, in the turning point $r=r_{t}$%
\begin{equation}
E_{c.m.}^{2}=2m^{2}+\frac{2E_{1}E_{2}}{N(r_{t})}.
\end{equation}%
By itself, $E_{c.m.}^{2}$ is finite. However, one can obtain a significant
energy gain as compared to collision in the flat space-time (\ref{minf})
even in this "trivial" scenario, provided both particles are
ultrarelativistic, $E_{1}\gg m,$ $E_{2}\gg m$. The corresponding additional
factor equals $\frac{E_{1}E_{2}}{m^{2}}$. As now $\frac{E_{1}}{m}$ and $%
\frac{E_{2}}{m}$ are free parameters, we can formally increase the energy
gain without a limit. The only difficulty is that we must have
ultrarelativistic particles from the very beginning. (To some extent, that
resembles the "energy feeding problem" discussed in Sec. IV C1 of \cite{axis}
for another scenario of collision in the extremal black hole background,
when particles move along the axis. Now, a similar problem reveals itself
for nonextremal ones and equatorial particle motion.)

\subsection{Near-extremal black holes}

Now, let us consider the limit $\kappa \rightarrow 0$. If $\kappa $ is small,%
\begin{equation}
N^{2}\approx 2\kappa (r-r_{+})+H(r-r_{+})^{2}\text{,}  \label{n}
\end{equation}%
where $H$ is the model-dependent coefficient. Then, the position of the
turning point $r_{t}$ for the critical particle is determined by equation $%
U=0.$ Taking into account (\ref{xrot}), we obtain from (\ref{U})%
\begin{equation}
(r_{t}-r_{+})C\approx 2\kappa (m^{2}+\frac{E^{2}}{\omega _{H}^{2}\left(
g_{\phi }\right) _{H}})\text{,}
\end{equation}%
where 
\begin{equation}
C=\frac{E^{2}}{\omega _{H}^{2}}(B_{1}^{2}-\frac{H}{\left( g_{\phi }\right)
_{H}})-Hm^{2}
\end{equation}%
and it is assumed that $C>0,$ subscript "H" refers to quantities calculated
on the horizon. Bearing in mind that $E\geq m$, it is sufficient to require
that $B_{1}^{2}>H(\omega _{H}^{2}+\frac{1}{\left( g_{\phi }\right) _{H}})$.

Then,%
\begin{equation}
r_{t}-r_{+}\approx \frac{2\kappa \tilde{m}^{2}(r_{+})}{C}\text{.}  \label{rc}
\end{equation}%
If $\kappa r_{+}\sim 1$ and $\frac{E}{m}\rightarrow \infty $, the numerator
has the same order as the denominator, so $r_{t}-r_{+}$ does not become
small in accordance with what is said after eq. (\ref{xrot}). However, for $%
\kappa \rightarrow 0$ we see that indeed $r_{t}\rightarrow r_{+}$.

For the critical particle 1, it follows from (\ref{xrot}) and (\ref{rc})
that 
\begin{equation}
X\approx \frac{E}{\omega _{H}}B_{1}\frac{2\kappa \tilde{m}^{2}(r_{+})}{C}%
\text{.}
\end{equation}%
Eq. (\ref{n}) gives us%
\begin{equation}
N^{2}(r_{t})\approx 4\frac{\kappa ^{2}\tilde{m}^{2}(r_{+})}{C}[1+H\frac{%
\tilde{m}^{2}(r_{+})}{C}]\text{.}
\end{equation}%
Then, it follows from (\ref{erot}) that%
\begin{equation}
E_{c.m.}^{2}\approx \frac{E_{1}\left( X_{2}\right) _{H}B_{1}C}{\kappa \omega
_{H}[C+H\tilde{m}^{2}(r_{+})]}\text{.}
\end{equation}%
Thus again%
\begin{equation}
E_{c.m.}^{2}\thicksim \frac{1}{\kappa }
\end{equation}%
can grow unbounded if $\kappa \rightarrow 0$.

For the Kerr metric,%
\begin{equation}
(\kappa r_{+})^{-1}\approx \frac{2}{\eta }\text{, }\eta =\sqrt{1-a_{\ast
}^{2}},
\end{equation}%
where $a_{\ast }=\frac{a}{M}$, $a$ being the standard parameter
characterizing an angular momentum, $M$ black hole mass.

In principle, there are three different scenarios: (i) the near-critical
particle is created already near the horizon \cite{gp}, (ii) collisions
involve a particle moving on a circular orbit near a black hole \cite%
{innermost}, (iii) both particles come from infinity and collide in the
turning point. Thus in all three cases there is only one small parameter
that is able to increase $E_{c.m.}^{2}$ significantly. It is the same for
all three scenarios and depends on the properties of a black hole. This is
the surface gravity $\kappa $ or, equivalently, $\eta $. If one takes the
astrophysically relevant limit $a_{\ast }=0,998$ \cite{thorne} one obtains
that $\eta ^{-1}\approx 22.361$. There are also numeric factors depending on
the scenario but we omit such details. We see that there exists enhancement
of the energy $E_{c.m.}^{2},$ although it remains bounded. Meanwhile, the
scenario under discussion gives some additional factors that can improve the
situation. It consists in the process that include superheavy particles. We
considered it in detail for the RN black hole but a similar phenomenon
should happen also for the Kerr one. Then, the ratio $\frac{m_{3}}{m}$ in
each additional collision can somehow increase the energy gain. Whether and
how this can be realized in a realistic astrophysical context is a separate
interesting question beyond the scope of the present paper.

\section{Discussion and conclusions}

Thus we considered two types of nonextremal black holes: charged static and
neutral rotating ones. In both cases, we considered scenarios in which the
critical and usual particles come from infinity and collide in the turning
point of the critical particle. Under some conditions, the location of this
point turns out to be close to the horizon. For the RN black hole, there are
two different factors that make it possible: either critical particle 1 is
ultralativistic or a black hole is near-extremal (or both factors are
valid). Then, $E_{c.m.}^{2}\thicksim \frac{E_{1}}{\kappa }$. On the first
glance, the necessity to have large $E_{1}$ from the very beginning,
depreciates the ability of a black hole to serve as a particle accelerator 
\cite{ks}. However, this is not so. One can compare, say, the scenario under
discussion to collision of two ultrarelativistic particles at flat infinity.
Then, we have significant gain in the energy of collisions if it happens
near the horizon. Also, for a moderate Killing energy $E_{1.2}\sim m$ the
energy of collision becomes indefinitely large if the surface gravity $%
\kappa $ is as small as we like. This is a counterpart of collisions on
near-circular orbits in the background of near-extremal black holes. There
exist two versions of the corresponding collisions in which $%
E_{c.m.}^{2}\thicksim \kappa ^{-1}$ similarly to our case or $%
E_{c.m.}^{2}\thicksim \kappa ^{-2/3}$ for two different types of scenarios 
\cite{innermost}, \cite{circ}. But now, the scenario has nothing to do with
the circular orbits, both particles come from infinity.

We also saw that if, after the first collision, new particle 3 collides
again with some particle that arrived from infinity, the energy $E_{c.m.}$
remains bounded in each individual collisions. However, if new created
near-critical particles are heavy enough with $m_{3}\gg m$, the process can
be repeated giving a growing factor proportional to $\left( m_{3}/m\right) $
for each new collision (where for simplicity we assumed that new
near-critical particles have the same mass $m_{3}$). Only an initial
particle with big $E_{1}$ is required, afterwards it is sufficient to send
from infinity particles with modest energy of the order $m.$

As far as the neutral rotating black hole is concerned, near-extremal black
holes with $\kappa \rightarrow 0$ are relevant in this context with the same
result $E_{c.m.}^{2}\thicksim \kappa ^{-1}$.

To summarize, there are two different types of accelerators connected with
black holes. The first type is presented by extremal black holes, where the
presence of the horizon reveals itself directly. It is the proximity of a
point of collision to the horizon (together with the fine-tuning of
parameters of one particle) that matters \cite{ban}, while the mass of
colliding particles are of secondary importance. Choosing this point close
enough to the horizon, one can obtain $E_{c.m.}$ as large as one likes
already in the first collision. For nonextremal black holes this is
impossible. But, nonetheless, nonextremal black hole can indeed be particle
accelerators, although with a number of restrictions described above. In
doing so, the relation between masses of particles that are created near the
horizon and those coming from infinity plays a crucial role in the scenario
of multiple collisions. It is able to enhance the initial gain in the energy
of collision. It would be interesting to consider more realistic
astrophysically relevant scenarios on the basis of the results obtained in
the present work.

\begin{acknowledgments}
The work is performed according to the Russian Government Program of
Competitive Growth of Kazan Federal University.
\end{acknowledgments}

\end{document}